\documentclass[english]{iopart}
\usepackage[T1]{fontenc}
\usepackage[latin1]{inputenc}
\usepackage{amssymb}

\makeatletter

\usepackage{iopams}

\usepackage{babel}
\makeatother

\begin{document}

\title[NMR probes for the Kondo scenario for the $0.7$ feature]{Nuclear magnetic resonance probes for the Kondo scenario for the
$0.7$ feature in semiconductor quantum point contact devices}

\author{V Tripathi$^{1}$ and N R Cooper$^{2}$}

\address{$^{1}$Department of Theoretical Physics, Tata Institute of Fundamental
Research, Homi Bhabha Road, Mumbai 400005, India}

\address{$^{2}$T. C. M. Group, Cavendish Laboratory, Department of Physics,
University of Cambridge, J. J. Thomson Avenue, Cambridge CB3 0HE,
United Kingdom}

\begin{abstract}
We discuss the expected features in nuclear relaxation and Knight
shift measurements for the Kondo scenario for the {}``$0.7$ feature''
in semiconductor quantum point contact (QPC) devices defined in two-dimensional
electron gases (2DEGs). As the conductance is more sensitive to the
nuclear polarisation in the centre of the QPC compared to that in
the 2DEG leads, our analysis is focused in the region near to the
centre of the QPC. We show that the exchange coupling of a bound electron
in the QPC with the nuclei would lead to, in the region near to the
centre of the QPC, a much higher rate of nuclear relaxation compared
to that involving exchange of nuclear spin with conduction electrons.
Away from the centre of the QPC, we find that the distance beyond
which the latter (conduction electron) mechanism becomes equally important
is of the order of typical QPC lengths; thus, between these two electronic
mechanisms, relaxation by coupling to the bound electron dominates
within the QPC. Furthermore, we show that the temperature dependence
of the nuclear relaxation due to coupling to the bound electron is
non-monotonic as opposed to the linear$-T$ relaxation from coupling
with conduction electrons. Nuclear spin diffusion processes restrict
the range of validity of this analysis. We present a qualitative analysis
of additional relaxation due to nuclear spin diffusion (NSD), and
compare the nuclear relaxation times associated with NSD and the above
electronic mechanisms. We discuss circumstances in which NSD will
affect our results significantly, and suggest ways in which NSD may
be suppressed in the QPC so that the Kondo physics may be unearthed.
Nuclear relaxation together with Knight shift measurements, will help
in verifying whether the {}``$0.7$'' feature is indeed due to the
presence of a bound electron in the QPC. While some of the results
have also been discussed in the context of paramagnetic impurities
in bulk conductors, our analysis is intended for application to the $0.7$
effect in semiconductor systems.
The qualitative and quantitative
estimates we make will allow experimental tests of the Kondo scenario for the
$0.7$ feature in QPCs in two-dimensional electron gas heterostructures.
\end{abstract}

\section{Introduction }

\subsection{The $0.7$ conductance anomaly}

The ballistic conductance $G$ of a quantum point contact (QPC) device,
measured as a function of the width of the channel transverse to the
current, is quantised in integer multiples of $G_{0}=2e^{2}/h$ in
the absence of a magnetic field and electron interactions. The application
of a strong in-plane magnetic field lifts the electron spin degeneracy
through Zeeman splitting without affecting the electron trajectories
in the plane of the device, and the quantisation then appears in multiples
of $G_{0}/2.$ These effects had been observed since 1988 \cite{wharam,vanWees},
and well-understood as arising from the quantisation of the electron
momentum in the QPC in the direction transverse to the current (transverse
sub-bands) \cite{vanWees,glazman}. A remarkable set of measurements
\cite{thomas-prl,thomas-prb-98,thomas-prb-2000}, beginning in 1996,
on the ubiquitous but hitherto overlooked additional {}``$0.7$ features''
between successive quantised plateaus of the ballistic conductance
has, since then, lead us to critically question our understanding
of electron transport in the humble QPC and directly inspired a great
deal of experimental \cite{kristensen-Vg-dep,reilly-interaction-evid,rokhinson-spinpol,cronenwett-kondo,abi-0.7-analogue,abi-moving-bands,rafi-nonlin-effects,appleyard-thermopower,roche-noise-red}
and theoretical \cite{wang-berggren,shelykh-spinpolar,havu-spinpolar,spivak-zhou-spinpol,klironomos-matveev-spinpol,flambaum-singlet-triplet,reilly-model,meir-kondo,cornaglia-kondo,bartosch-fm-luttinger,shushkov-cdw,reimann-spin-peierls,matveev-wigner,seelig-matveev}
work.

Some of the salient features of this {}``$0.7$ effect,'' as it
is usually referred to are as follows. The ballistic conductance,
as a function of the gate voltage (that controls the cross-sectional
width of the QPC), shows shoulder-like structures at the {}``steps''
marking the transitions between successive quantised conductance plateaus,
$G_{n}=nG_{0}.$ The shoulders usually occur at values of around $0.7G_{n}$
between neighbouring quantised plateaus $G_{n}$ and $G_{n+1},$ \cite{thomas-prl,thomas-prb-98}
although their positions are not universal and have been known to
occur as low as $0.5G_{0}$ \cite{thomas-prb-2000,reilly-interaction-evid}.
The shoulders are not due to disorder effects nor they are transmission
resonances \cite{thomas-prb-98}. The most prominent shoulder occurs
where the QPC makes a transition from a pinch-off state ($n=0$) to
the first quantised plateau. The temperature dependence of this feature
is very unusual \cite{thomas-prl,thomas-prb-98,kristensen-Vg-dep,reilly-interaction-evid,cronenwett-kondo}.
Decreasing the temperature makes it less well-defined, and it altogether
disappears at low temperatures of the order of a few tens of millikelvins.
Increasing the temperature makes the feature more well-defined, until,
beyond a few kelvins, the feature as well as the quantised plateaus
begin to get thermally smeared out. The temperature dependence has
been fitted with an Arrhenius law \cite{kristensen-Vg-dep} as well
as a power law \cite{cronenwett-kondo}, and the conductance change
over the temperature range in which the feature exists is insufficient
to resolve this ambiguity. The characteristic temperature scale associated
with the feature is of the order of a kelvin. Upon the application
of an in-plane magnetic field that removes electron spin degeneracy
through Zeeman splitting without affecting their trajectories in the
plane of the device, the $0.7$ shoulder shifts lower in a smooth
manner, finally moving to $0.5G_{0}$ at fields of the order of a
few tesla (corresponding to complete lifting of electron spin degeneracy).
This is evidence that the feature is intimately connected with electron
spin. The $0.7$ feature is believed to arise due to electron interaction
\cite{thomas-prl,thomas-prb-98,kristensen-Vg-dep,reilly-interaction-evid,cronenwett-kondo}
as can be seen from the following two characteristic features. The
gyromagnetic ratio $g_{e}$ of the electrons is larger in the lowest
sub-bands by a factor of about two compared with the bulk GaAs value
of $g_{e}=-0.44,$ and decreases towards $-0.44$ in the higher sub-bands
\cite{thomas-prl,thomas-prb-98}. Enhancement of the gyromagnetic
ratio may be associated with electron interaction. Since in the lower
sub-bands, the number of electrons in the QPC is smaller and electrostatic
screening is weaker, electron interaction effects such as exchange
are expected to be stronger there. In presence of a non-zero source-drain
potential difference $V_{sd},$ $dG/dV_{sd}$ shows a zero-bias anomaly
(peak) at $V_{sd}=0,$ which is not generally expected for noninteracting
electrons \cite{cronenwett-kondo}.

Numerous scenarios have been studied for the $0.7$ feature ranging
from electron spin polarisation in the QPC \cite{thomas-prl,kristensen-Vg-dep,reilly-interaction-evid,rokhinson-spinpol,wang-berggren,shelykh-spinpolar,havu-spinpolar,spivak-zhou-spinpol,klironomos-matveev-spinpol,reilly-model},
exchange splitting of few electron bound states in the QPC \cite{flambaum-singlet-triplet},
Kondo effect arising from quasi-bound electrons in the QPC \cite{cronenwett-kondo,meir-kondo,cornaglia-kondo},
ferromagnetic Luttinger liquids \cite{bartosch-fm-luttinger}, charge
\cite{shushkov-cdw} and spin density waves\cite{reimann-spin-peierls},
and Wigner crystallisation effects in one dimension \cite{klironomos-matveev-spinpol,matveev-wigner}.
Of these, the electron spin polarisation and Kondo scenarios have
been most extensively studied, while the Wigner crystallisation scenario
is a more recent proposal that also looks promising.

Choosing theoretically between the electron spin polarisation and
Kondo pictures has proved difficult because both have been able to
substantially describe the experimental observations. Recent measurements
of the $0.7$ feature in hole-doped GaAs in Ref. \cite{rokhinson-spinpol}
used two QPCs in a hole-focusing setup that claimed to confirm the
spin polarisation picture and rule out the Kondo picture as incompatible
with their data. On the other hand, features such as the zero bias
anomaly observed in measurements at non-zero $V_{sd}$ \cite{cronenwett-kondo}
have not been explained using the spin polarisation picture, although
there has been a suggestion that the zero bias anomaly can also arise
from backscattering by acoustic phonons \cite{seelig-matveev}.

\subsection{NMR for the $0.7$ feature}

In this paper we discuss the signatures in nuclear relaxation of the
presence of a bound electron in a short QPC. We present a fairly detailed
review on nuclear relaxation in the presence of a bound electron in
the QPC. The purpose is twofold. First, these NMR methods are not
yet being used in the $0.7$ community and an analysis of nuclear
relaxation in this context may be useful. Second, we have recently
studied \cite{cooper07} nuclear relaxation in QPCs for the Kondo
scenario as well as for the other proposed physical mechanisms for
the $0.7$ feature. Here we present details of the calculations for
the {}``Kondo'' part in Ref. \cite{cooper07}, and also discuss
in addition, the effects of nuclear spin diffusion processes on the
relevance of the analysis. 

Nuclear relaxation measurements in nanoscale systems such as QPCs
have been hampered, in comparison with bulk systems, by the small
number of polarised nuclei. Recently, however, it has been shown how
nuclear polarisation may be created \cite{tripathi07,cooper07} and
detected \cite{cooper07,nesteroff} in QPCs through the measurement
of the two-terminal conductance. In this paper, we devote our attention
to the region near the centre of the QPC as the conductance is more
sensitive to nuclear polarisation in this region than it is to nuclear
polarisation away from the QPC in the 2DEG leads. 

We compare the nuclear relaxation rates from the coupling of the nuclei
with (a) the bound electron and (b) the conduction electrons both
above and below the Kondo temperature $T_{K}.$ We show that near
to the centre of the QPC, the relaxation through coupling with the
bound electron will be in general much faster, and furthermore, follow
a (very different) non-monotonous temperature dependence. In the high temperature
regime ($T > T_K$) the relaxation rates, respectively, due to impurity 
coupling, $T_{1}^{imp}$, and conduction electrons, $T_{1}^{cond-el}$, are given by (see Eq.(\ref{eq:T-par-perp-highT}) and Eq.(\ref{eq:cond-el-highT}))
\begin{eqnarray}
\frac{1}{T_{1}^{imp}} & = & \frac{2A_{d}(\mathbf{R}_{i})^{2}S(S+1)}{3\pi\hbar(k_{B}T)(J\rho(\epsilon_{F}))^{2}}, \\
\frac{1}{T_{1}^{cond-el}} & = & \frac{\pi(k_{B}T)}{\hbar}(A_{s}\rho(\epsilon_{F}))^{2}.
\qquad\qquad\mbox{(high temp.)}
\end{eqnarray}
Here $A_{d}(\mathbf{R}_{i})$ is the hyperfine interaction of the nucleus at point $\mathbf{R}_{i}$ with the impurity spin $S=1/2$ at the origin, $A_s$ is the hyperfine interaction of the nucleus with the conduction electrons and $J$ is the interaction of the impurity spin and conduction electrons. $\rho(\epsilon_{F})$ is the density of states at the Fermi energy. In the low temperature regime ($T<T_K$), the relaxation rates in the two cases are (see Eq.(\ref{eq:T-par-perp-kondo}) and Eq.(\ref{eq:enhanced-rho-T1}))
\begin{eqnarray}
\frac{1}{T_{1}^{imp}} & = & \frac{4\pi(k_{B}T)A_{d}(\mathbf{R}_{i})^{2}}{\hbar(g_{s}\mu_{B})^{4}}\chi_{\mbox{imp}}^{2}, \\
\frac{1}{T_{1}^{cond-el}} & = & \frac{\pi(k_{B}T)}{\hbar}(A_{s}\rho(\epsilon_{F}))^{2}\left(1+\frac{2C}{N}\frac{T_F}{T_K}\right). \qquad\mbox{(low temp.)}
\end{eqnarray}
Here $\chi_{\mbox{imp}}$ is the susceptibility of the impurity spin, $C$ is a constant of the order one, $T_F$ is the Fermi temperature and $N$ is the number of electrons in the QPC. The conduction electron results at low temperatures and high temperatures differ only through the enhancement of the density of states of the conduction electrons that occurs below the Kondo temperature. For details of these results we refer the reader to Sec. \ref{sec:Temperatures-above-T_{K}}  and Sec. \ref{sec:Temperatures-below-T_{K}}. 
Associating
the (experimentally observed) characteristic temperature scale $\sim1{\rm K}$
associated with the $0.7$ feature with $T_{K},$ we have the following
estimates. For $T=2K$ (high temperature regime), the nuclear relaxation
times associated with processes (a) and (b) near the centre of the
QPC are respectively $T_{1}^{imp}\approx0.1{\rm s}$ and $T_{1}^{cond-el}\approx5{\rm s}.$
For $T=0.5{\rm K}$ (low temperature regime), we find $T_{1}^{imp}\approx3.5\times10^{-2}{\rm s}$
and $T_{1}^{cond-el}\approx20{\rm s}.$ Away from the centre of the
QPC, the nuclear relaxation rate due to impurity coupling decreases
as the exchange (RKKY) interaction of the bound electron and a nuclear
spin at a distance $R_{i}$ from the electron falls off as $1/(k_{F}R_{i}).$
We show in Sec. \ref{sec:Discussion1} that below the Kondo temperature, relaxation by coupling to
conduction electrons dominates at distances beyond $R_{i}=(4\epsilon_{F}/k_{B}T_{K}k_{F}),$
where $\epsilon_{F}$ is the Fermi energy of the electrons in the
QPC. For a 2D electron density of $10^{11}{\rm cm}^{-2},$ 1D Fermi
energy of $20{\rm K},$ and a Kondo temperature of $1{\rm K},$ we
estimate this distance $R_{i}$ to be about $1.6\mu{\rm m},$ which
is of the order of the length of typical QPCs. Since nuclear relaxation
in the QPC affects the conductance far more than that in the 2DEG
leads, we thus conclude that between these two electronic mechanisms,
the conductance is determined more by the nuclear relaxation from
coupling to the impurity electron than by coupling to the conduction
electrons. 

The final test for a bound electron, which we propose here, comes
from Knight shift measurements. The temperature dependence of the
Knight shift is shown to be the same as the temperature dependence
of the susceptibility of a Kondo impurity. The Knight shift may be
measured by observing the conductance as a function of the frequency
of an external electromagnetic wave to which the QPC is subjected.
When the frequency matches the difference in energy of successive
nuclear Zeeman levels, the nuclear polarisation will get destroyed
resulting in a sudden change in conductance. 

Internuclear dipolar interactions give rise to non-conserving spin
flips and internuclear flip-flops, and limit the range of validity
of our analysis. In GaAs, these interactions correspond to a field
of the order of a millitesla which is equivalent to $T_{1}\sim T_{2}\sim10^{-4}{\rm s}$
in the absence of a magnetic field. However in a non-zero magnetic field
of several millitesla, this intrinsic $T_{1}$ may be many orders
of magnitude larger (see Sec. \ref{sec:discussion2}); therefore the
measurements we propose should be performed in the presence of small
but non-zero magnetic fields. Apart from non-conserving spin flips,
internuclear spin flip-flop processes can be significant
even in the presence of a magnetic field, and cause nuclear spin diffusion
(NSD). Our most conservative estimate (see Sec. \ref{sec:discussion2})
for the nuclear spin diffusion time for the QPC is $T_{1}^{sd}\sim0.4{\rm s}$
which is based on using the bulk value for the nuclear spin diffusion
constant in GaAs. However, as we discuss later, the nuclear spin diffusion
constant for a QPC with a localised electron can be much smaller than
the bulk value because the resulting non-uniformity of the hyperfine
interaction suppresses internuclear flip-flops. We review recent literature
on NSD in quantum dots where it has been shown that NSD can be further
suppressed by one to two orders of magnitude by applying fields greater
than $1{\rm mT},$ and also by suitable redesigning of the heterostructure
as for example by growing AlGaAs layers on either side of the GaAs
layer. We believe that the fairly long relaxation times associated
with NSD in QPCs (or quantum dots) together with the possibility of
further strong suppression of NSD through small magnetic fields and/or
device redesigning makes it quite feasible to observe nuclear relaxation
effects due to the bound electron in the QPC.

The nuclear relaxation and Knight shift measurements together enable
a confirmation of the presence of a bound electron in the QPC, if
any.

The rest of the paper is organised as follows. We introduce our model
in Sec. \ref{sec:Model} for a QPC with a bound electron and provide
general expressions of the experimentally measured nuclear relaxation
rates $T_{1}^{-1}$ and $T_{2}^{-1}.$ In Sec. \ref{sec:Temperatures-above-T_{K}}
and Sec. \ref{sec:Temperatures-below-T_{K}}, respectively, we analyze
the nuclear relaxation at temperatures above and below the Kondo temperature.
The crossover between the high and low temperature regimes is discussed
in Sec. \ref{sec:Crossover}, and Sec. \ref{sec:Discussion1} contains
a discussion of the relative strengths of nuclear relaxation by coupling
to conduction electrons and by coupling to the bound electron spin.
Finally in Sec. \ref{sec:discussion2}, we discuss nuclear spin diffusion
(NSD) effects, how it affects our earlier analysis, and ways in which
NSD can be suppressed so that the Kondo scenario for the $0.7$ feature
may be feasibly tested with the proposed NMR method.

\section{Model \label{sec:Model}}

We consider a simple model of a QPC defined in a two-dimensional electron
gas (2DEG) in the $xz$ plane, taking the transport direction along
the $x$ axis. Let $w_{x},$ $w_{z}$ be the dimensions of the QPC
in the $xz$ plane, and $w_{y}$ in the direction perpendicular to
the 2DEG. We assume the bound electron (impurity) of spin $\mathbf{S}$
is localised at the origin $\mathbf{r}=0$ which we take as the centre
of the QPC. Let $\mathbf{I}_{i}$ be the nuclear spins of the host
GaAs, and the conduction electron spin density be denoted by $\boldsymbol{\sigma}(\mathbf{r}).$
The Hamiltonian is \begin{eqnarray}
H & = & \sum_{\mathbf{k}\boldsymbol{\sigma}}\epsilon_{\mathbf{k}}c_{\mathbf{k}\boldsymbol{\sigma}}^{\dagger}c_{\mathbf{k}\boldsymbol{\sigma}}-\mathbf{H}_{0}\cdot\left(g_{s}\mu_{B}\mathbf{S}+g_{n}\mu_{n}\sum_{i}\mathbf{I}_{i}+g_{\sigma}\mu_{B}\sum_{i}\boldsymbol{\sigma}(\mathbf{r}_{i})\right)+\nonumber \\
 & \qquad & +J\mathbf{S}\cdot\boldsymbol{\sigma}(0)+A_{s}\sum_{i}\mathbf{I}_{i}\cdot\boldsymbol{\sigma}(\mathbf{R}_{i})+A_{d}\mathbf{I}_{0}\cdot\mathbf{S}.\label{eq:hamilt1}\end{eqnarray}
 $\mathbf{H}_{0}$ is the external magnetic field. We assume that
the electronic Zeeman energy $g_{\sigma}\mu_{B}\sum_{i}\mathbf{H}_{0}\cdot\boldsymbol{\sigma}(\mathbf{r}_{i})$
is much less than the Kondo temperature associated with the (antiferromagnetic)
impurity-conduction electron coupling $J,\,(J>0),$ such that the
Kondo is not suppressed by Zeeman splitting. $A_{s}$ is the hyperfine
coupling strength between the nuclei and conduction electrons. It
is of the order of $100\mu eV$ per nucleus in GaAs. The hyperfine
contact term $A_{d}$ coupling the impurity electron to the nuclear
spins is proportional to the probability density of the localised
electron wavefunction at the origin. Near to the centre of the QPC,
\begin{eqnarray}
A_{d} & \approx & \frac{8A_{s}}{(w_{x}w_{y}w_{z})}.\label{eq:Ad-local}\end{eqnarray}
 The impurity spin is localised over a volume, typically, $w_{x}w_{y}w_{z}\sim1\mu{\rm {m}\times5{\rm {nm}\times20{\rm {nm}}}}$,
that greatly exceeds the volume per nucleus $\sim1{\rm {nm}^{3}.}$
At temperatures much lower than the Fermi temperature, we may assume
the impurity electron remains in the lowest energy state of the potential
confining it. In the absence of the coupling of the impurity spin
to the conduction electrons, the impurity susceptibility would have
obeyed the Curie law. At temperatures small compared to the Fermi
temperature, this susceptibility would be larger than the corresponding
Pauli susceptibility of the conduction electrons.

We ignore the direct magnetic dipolar interaction of the nuclear spins.
In the volume $V_{0}=w_{x}w_{y}w_{z}$ where the impurity electron
is localised, we will show that the contribution to nuclear relaxation
from the coupling of the nuclear spin with the conduction electrons
would be small compared to the contribution from the nuclear coupling
with the localised electron. The reason is that the localised electron
corresponds to an enhanced spin density compared to the conduction
electrons. We can also ignore the indirect exchange (RKKY) interaction
of different nuclei as its strength would be small, of the order of
$A_{s}^{2}.$ However it is important to retain the RKKY interaction
of the localised electron with distant nuclei, especially those lying
outside $V_{0}.$ The strength of this interaction is proportional
to $JA_{s}\gg A_{s}^{2}$ (electronic energy scales such as $J$ are
expected to be typically larger than corresponding nuclear energy
scales such as $A_{s}$). The RKKY hyperfine interaction will be of
the form \begin{eqnarray}
H_{RKKY}(\mathbf{R}_{i}) & = & A_{RKKY}(\mathbf{R}_{i})\mathbf{I}_{i}\cdot\mathbf{S},\label{eq:H-RKKY}\end{eqnarray}
 where, for $k_{F}R_{i}\gg1$ and one spatial dimension, the RKKY
interaction is \cite{yafet}\begin{eqnarray}
A_{RKKY}(\mathbf{R}_{i}) & \approx & -\frac{JA_{s}\rho(\epsilon_{F})}{V_{0}}\left[\frac{\pi}{2}-\mbox{Si}(2k_{F}R_{i})\right],\label{eq:A-RKKY}\end{eqnarray}
 where $\rho(\epsilon_{F})=4m/(2\pi\hbar^{2}k_{F}w_{y}w_{z})$ is
the density of electron states in the QPC and $\mbox{Si}(x)$ is the
sine integral function. At large values of its argument, $\mbox{Si}(x)\approx\pi/2-\cos(x)/x-\sin(x)/x^{2},$
while for small values of $x,$ $\mbox{Si}(x)\approx x.$ The hyperfine
interaction $A_{d}$ for the nuclei near the centre of the QPC (given
by (\ref{eq:Ad-local})) as well as $A_{RKKY}(\mathbf{R}_{i})$ for
those further away can be conveniently expressed by introducing a
spatially varying hyperfine coupling $A_{d}(\mathbf{R}_{i}):$ \begin{eqnarray}
H_{I,S} & = & \sum_{i}A_{d}(\mathbf{R}_{i})\mathbf{I}_{i}\cdot\mathbf{S}.\label{eq:H-IS}\end{eqnarray}

The coupling of a nuclear spin with its external environment can be
written as \begin{eqnarray}
H_{n}(\mathbf{R}_{i}) & = & -g_{n}\mu_{n}(\mathbf{H}_{0}+\mathbf{H}_{\mbox{loc}}(\mathbf{R}_{i}))\cdot\mathbf{I}_{i},\label{eq:eff-mag-field}\end{eqnarray}
 where \begin{eqnarray}
\mathbf{H}_{\mbox{loc}}(\mathbf{R}_{i}) & = & -\frac{1}{g_{n}\mu_{n}}(A_{s}\boldsymbol{\sigma}(\mathbf{R}_{i})+A_{d}(\mathbf{R}_{i})\mathbf{S})\label{eq:H-loc}\end{eqnarray}
 is the local field due to electrons at the site $\mathbf{R}_{i}.$
The second contribution in (\ref{eq:H-loc}) is more important when
the impurity to host nucleus distance is not very large because the
susceptibility of the localised spin, $\sim\mu_{B}^{2}/k_{B}T$ is
a factor $\epsilon_{F}/k_{B}T$ larger than the Pauli susceptibility
per conduction electron.

The local field is the sum of an {}``average'' part $\langle\mathbf{H}_{\mbox{loc}}\rangle$
and a fluctuation part $\delta\mathbf{H}_{\mbox{loc}}.$ The nuclear
resonance occurs at a frequency $\omega_{n}$ given by \begin{eqnarray*}
\hbar\omega_{n}(\mathbf{R}_{i}) & = & g_{n}\mu_{n}H_{0}(1+K(\mathbf{R}_{i})),\end{eqnarray*}
 where \begin{eqnarray}
K(\mathbf{R}_{i}) & = & \langle H_{\mbox{loc}}^{z}(\mathbf{R}_{i})\rangle/H_{0}\label{eq:knight-shift-def}\end{eqnarray}
 is the Knight shift. The Knight shift in general depends on the location
$\mathbf{R}_{i}.$

The longitudinal and transverse nuclear relaxation rates (due to local
field fluctuations) $T_{\parallel}^{-1}$ and $T_{\perp}^{-1}$ are
respectively \cite{narath}\begin{eqnarray}
T_{\parallel}^{-1}(\mathbf{R}_{i}) & = & \frac{(g_{n}\mu_{n})^{2}}{2\hbar^{2}}\int_{-\infty}^{\infty}dt\,\langle\delta H_{\mbox{loc}}^{z}(\mathbf{R}_{i},t)\delta H_{\mbox{loc}}^{z}(\mathbf{R}_{i},0)\rangle,\nonumber \\
T_{\perp}^{-1}(\mathbf{R}_{i}) & = & \frac{(g_{n}\mu_{n})^{2}}{4\hbar^{2}}\int_{-\infty}^{\infty}dt\, e^{i\omega_{n}t}\langle\delta H_{\mbox{loc}}^{+}(\mathbf{R}_{i},t)\delta H_{\mbox{loc}}^{-}(\mathbf{R}_{i},0)\rangle.\label{eq:t-parallel-perp}\end{eqnarray}
 These are related to the \emph{experimentally measured} longitudinal
relaxation rate $T_{1}^{-1}$ and transverse relaxation rate $T_{2}^{-1}$
through \cite{narath}\begin{eqnarray}
T_{1}^{-1} & = & 2T_{\perp}^{-1},\nonumber \\
T_{2}^{-1} & = & T_{\parallel}^{-1}+T_{\perp}^{-1}.\label{eq:t1-t2-def}\end{eqnarray}
 Thus the Knight shifts as well as the nuclear relaxation rates depend
on the locations of the nuclei.

It is possible to express the correlators of the fluctuating magnetic
fields in (\ref{eq:t-parallel-perp}) in terms of the dynamic susceptibility
$\chi^{\alpha\beta}(\mathbf{R}_{i},\omega)$ using the fluctuation-dissipation
theorem. Here $\alpha$ and $\beta$ are the longitudinal $(z)$ and
transverse $(+,-)$ labels. The fluctuation-dissipation theorem gives\begin{eqnarray}
\mbox{Im}\chi^{\alpha\beta}(\mathbf{R}_{i},\omega) & = & \frac{1}{\hbar}\tanh\left(\frac{\hbar\omega}{2k_{B}T}\right)C^{\alpha\beta}(\mathbf{R}_{i},\omega),\label{eq:fluc-diss-theorem}\end{eqnarray}
 where \begin{eqnarray}
C^{\alpha\beta}(\mathbf{R}_{i},\omega) & = & \int_{-\infty}^{\infty}dt\, e^{i\omega t}\langle\delta M^{\alpha}(\mathbf{R}_{i},t)\delta M^{\beta}(\mathbf{R}_{i},0)\rangle\label{eq:mag-correlator}\end{eqnarray}
 is the correlator of the fluctuations of the magnetic moment $\mathbf{M}.$
At low frequencies $\omega\ll k_{B}T/\hbar,$ (\ref{eq:fluc-diss-theorem})
simplifies to \begin{eqnarray*}
\mbox{Im}\frac{\chi^{\alpha\beta}(\mathbf{R}_{i},\omega)}{\omega} & \approx & \left(\frac{\omega}{2k_{B}T}\right)C^{\alpha\beta}(\mathbf{R}_{i},\omega).\end{eqnarray*}

We now study two extreme cases. The first concerns nuclei not very
far from the impurity so that the relaxation of the nuclei is dominated
by their coupling to the impurity. The second case concerns distant
nuclei where the RKKY interaction is small and the nuclear relaxation
is dominated by their coupling to the conduction electrons. We will
study the nuclear relaxation both above and below the Kondo temperature
of the impurity electron.

\section{Temperatures above $T_{K}$ \label{sec:Temperatures-above-T_{K}}}

\subsection{Relaxation due to impurity coupling}

The local field at a nucleus at $\mathbf{R}_{i}$ has a simple relation
with the magnetic moment $\mathbf{M}$ of the impurity electron: \begin{eqnarray}
\mathbf{H}_{\mbox{loc}}(\mathbf{R}_{i}) & \approx & -\frac{A_{d}(\mathbf{R}_{i})}{g_{n}\mu_{n}}\mathbf{S}=-\frac{A_{d}(\mathbf{R}_{i})}{g_{n}g_{s}\mu_{n}\mu_{s}}\mathbf{M}.\label{eq:Hloc-M-relation}\end{eqnarray}
 Using this relation between $\mathbf{M}$ and $\mathbf{H}_{\mbox{loc}}$
together with (\ref{eq:t-parallel-perp}) and (\ref{eq:fluc-diss-theorem}),
the nuclear relaxation rates at low frequencies can be shown to be
\begin{eqnarray}
T_{\parallel}^{-1}(\mathbf{R}_{i}) & = & k_{B}T\left(\frac{A_{d}(\mathbf{R}_{i})}{\hbar g_{s}\mu_{B}}\right)^{2}\mbox{Im}\frac{\chi_{\mbox{imp}}^{zz}(\omega)}{\omega}\bigg|_{\omega\rightarrow0},\label{eq:T-parallel-IS}\end{eqnarray}
 and \begin{eqnarray}
T_{\perp}^{-1} &  & (\mathbf{R}_{i})=\frac{1}{4\hbar}\left(\frac{A_{d}(\mathbf{R}_{i})}{g_{s}\mu_{B}}\right)^{2}\coth\left(\frac{\hbar\omega_{n}}{2k_{B}T}\right)\mbox{Im}\chi_{\mbox{imp}}^{+-}(\omega_{n}).\label{eq:T-perp-IS}\end{eqnarray}
 In our case, $k_{B}T$ is much larger than the nuclear Zeeman energy
$\hbar\omega_{n},$ so $T_{\perp}^{-1}$ is approximately \begin{eqnarray}
T_{\perp}^{-1} & ( & \mathbf{R}_{i})=\frac{k_{B}T}{2}\left(\frac{A_{d}(\mathbf{R}_{i})}{\hbar g_{s}\mu_{B}}\right)^{2}\mbox{Im}\frac{\chi_{\mbox{imp}}^{+-}(\omega)}{\omega}\bigg|_{\omega\rightarrow0}.\label{eq:T-perp-IS2}\end{eqnarray}
 $\chi_{\mbox{imp}}$ is the susceptibility of the impurity electron.
We need to obtain expressions for the imaginary part of the impurity
susceptibility.

Let $T_{e1}^{-1}$ and $T_{e2}^{-1}$ be the longitudinal and transverse
relaxation times for the impurity, and let $\chi_{\mbox{imp}}^{L}$
and $\chi_{\mbox{imp}}^{T}$ be respectively the longitudinal and
transverse static impurity susceptibilities: \begin{eqnarray}
\chi_{\mbox{imp}}^{L} & = & g_{s}\mu_{B}\partial\langle S_{z}\rangle/\partial H_{0},\nonumber \\
\chi_{\mbox{imp}}^{T} & = & g_{s}\mu_{B}\langle S_{z}\rangle/H_{0}.\label{eq:chi-L-T}\end{eqnarray}
 At small magnetic fields, there is no difference between the static
longitudinal and transverse impurity susceptibility. Expressions for
the imaginary part of the impurity susceptibility are available in
the literature \cite{gotze}:\begin{eqnarray}
\mbox{Im}\frac{\chi_{\mbox{imp}}^{zz}(\omega)}{\omega} & = & \chi_{\mbox{imp}}^{L}\frac{T_{e1}}{1+(\omega T_{e1})^{2}},\nonumber \\
\mbox{Im}\frac{\chi_{\mbox{imp}}^{+-}(\omega)}{2\omega} & = & \chi_{\mbox{imp}}^{T}\frac{T_{e2}}{1+[(\omega-\omega_{e})T_{e2}]^{2}}.\label{eq:Im-chi}\end{eqnarray}
 $T_{e1}$ and $T_{e2}$ also depend on the frequency but we are interested
only in the zero frequency limits. From Ref. \cite{narath}, \begin{eqnarray}
T_{e1}^{-1} & = & T_{e2}^{-1}=\frac{\pi k_{B}T}{\hbar}(J\rho(\epsilon_{F}))^{2},\quad\omega_{e}T_{e2}\ll1,\label{eq:Te-high-temp}\\
T_{e1}^{-1} & = & 2T_{e2}^{-1}=\pi S(J\rho(\epsilon_{F}))^{2}\omega_{e},\quad\omega_{e}T_{e2}\gg1.\label{eq:Te-low-temp}\end{eqnarray}
 The corresponding expressions for the imaginary part of the impurity
susceptibility may obtained from (\ref{eq:Im-chi}) by substituting
the values of the static transverse and longitudinal impurity susceptibility
defined in (\ref{eq:chi-L-T}).\cite{narath} For $\omega_{e}T_{e2}\ll1$
we have \begin{eqnarray}
\mbox{Im}\frac{\chi_{\mbox{imp}}^{zz}(\omega)}{\omega}\bigg|_{\omega\rightarrow0}=\mbox{Im}\frac{\chi_{\mbox{imp}}^{+-}(\omega)}{2\omega}\bigg|_{\omega\rightarrow0}\qquad\nonumber \\
=\frac{2\hbar S(S+1)(g_{s}\mu_{B})^{2}}{3\pi(k_{B}T)^{2}(J\rho(\epsilon_{F}))^{2}},\quad\omega_{e}T_{e2}\ll1;\label{eq:im-chi-highT}\end{eqnarray}
 thus the nuclear relaxation rates are\begin{eqnarray}
T_{\parallel}^{-1}(\mathbf{R}_{i})=T_{\perp}^{-1}(\mathbf{R}_{i})=\frac{A_{d}(\mathbf{R}_{i})^{2}S(S+1)}{3\pi\hbar(k_{B}T)(J\rho(\epsilon_{F}))^{2}},\quad\omega_{e}T_{e2}\ll1.\label{eq:T-par-perp-highT}\end{eqnarray}
 For $\omega_{e}T_{e2}\gg1,$ which is the case at low temperatures
and/or high fields, the fluctuations are very anisotropic. The imaginary
part of the impurity susceptibility and the corresponding nuclear
relaxation rates are \begin{eqnarray}
\mbox{Im}\frac{\chi_{\mbox{imp}}^{+-}(\omega)}{2\omega}\bigg|_{\omega\rightarrow0} & = & \frac{\pi S^{2}(g_{s}\mu_{B})^{2}(J\rho(\epsilon_{F}))^{2}}{2\omega_{e}^{2}},\nonumber \\
T_{\perp}^{-1}(\mathbf{R}_{i}) & = & \frac{\pi(k_{B}T)A_{d}(\mathbf{R}_{i})^{2}S^{2}(J\rho(\epsilon_{F}))^{2}}{2\hbar^{3}\omega_{e}^{2}},\label{eq:T-perp-lowT}\end{eqnarray}
 and \begin{eqnarray}
\mbox{Im}\frac{\chi_{\mbox{imp}}^{zz}(\omega)}{\omega}\bigg|_{\omega\rightarrow0} & = & \frac{(g_{s}\mu_{B})^{2}e^{-\hbar\omega_{e}/k_{B}T}}{\pi\omega_{e}S(k_{B}T)(J\rho(\epsilon_{F}))^{2}},\nonumber \\
T_{\parallel}^{-1}(\mathbf{R}_{i}) & = & \frac{A_{d}(\mathbf{R}_{i})^{2}e^{-\hbar\omega_{e}/k_{B}T}}{\pi\hbar^{2}\omega_{e}(J\rho(\epsilon_{F}))^{2}}\approx0,\quad\omega_{e}T_{e2}\gg1.\label{eq:T-par-lowT}\end{eqnarray}
 The experimentally observed relaxation rates $T_{1}^{-1}$ and $T_{2}^{-1}$
are obtained by using the relations in (\ref{eq:t1-t2-def}).

\subsection{Relaxation due to conduction electron coupling}

Expressions for nuclear relaxation due to coupling to conduction electrons
can be obtained by substituting $J,$ $\omega_{e}$ and $T_{e}$ in
(\ref{eq:Te-high-temp}) by $A_{s},$ $\omega_{n}$ and $T_{n}.$
Since the nuclear Zeeman energy is so small, we will always be interested
in the high temperature case. The result is \cite{korringa}\begin{eqnarray}
T_{1}^{-1}=T_{2}^{-1} & = & \frac{\pi(k_{B}T)}{\hbar}(A_{s}\rho(\epsilon_{F}))^{2}.\label{eq:cond-el-highT}\end{eqnarray}
 Note that the nuclear relaxation due to coupling to the impurity
spin does not have a Korringa-like temperature dependence. This may
be regarded as a signature of the presence of a localised electron.

\section{Temperatures below $T_{K}$ \label{sec:Temperatures-below-T_{K}}}

Let us now consider nuclear relaxation below the Kondo temperature
$T_{K}.$ The more interesting case, again, is that of relaxation
by coupling to the impurity spin.

\subsection{Relaxation due to impurity spin}

The following analysis presumes that $g_{s}\mu_{B}H_{0}/k_{B}T_{K}\ll1.$
For higher fields, the analysis of Sec. \ref{sec:Temperatures-above-T_{K}}
should be used. At small fields, we have mentioned earlier that there
is no difference between the static longitudinal and transverse susceptibilities.
When $T\ll T_{K},$ the imaginary part of the susceptibility satisfies
an elegant relation, \cite{shiba}

\begin{eqnarray}
\mbox{Im}\frac{\chi_{\mbox{imp}}^{zz}(\omega)}{\omega}\bigg|_{\omega\rightarrow0}=\mbox{Im}\frac{\chi_{\mbox{imp}}^{+-}(\omega)}{2\omega}\bigg|_{\omega\rightarrow0} & = & \frac{2\pi\hbar\chi_{\mbox{imp}}^{2}}{(g_{s}\mu_{B})^{2}}.\label{eq:shiba}\end{eqnarray}
 As a result, the nuclear relaxation rates take the simple form\begin{eqnarray}
T_{\parallel}^{-1}=T_{\perp}^{-1} & = & \frac{2\pi(k_{B}T)A_{d}(\mathbf{R}_{i})^{2}}{\hbar(g_{s}\mu_{B})^{4}}\chi_{\mbox{imp}}^{2}.\label{eq:T-par-perp-kondo}\end{eqnarray}

Using (\ref{eq:Hloc-M-relation}) in the definition of the Knight
shift, (\ref{eq:knight-shift-def}), it is easy to see that \begin{eqnarray}
K(\mathbf{R}_{i}) & = & \frac{A_{d}(\mathbf{R}_{i})\mbox{Re}\chi_{\mbox{imp}}^{zz}(0)}{(g_{n}\mu_{n})(g_{s}\mu_{B})}.\label{eq:knight-shift2}\end{eqnarray}
 $\mbox{Re}\chi^{zz}(0)$ is just the static impurity susceptibility
$\chi_{\mbox{imp}}.$ Eq.(\ref{eq:knight-shift2}) is also valid
above the Kondo temperature. Different nuclei will couple with the
impurity with different strengths $A_{d}(\mathbf{R}_{i});$ however,
the temperature dependence of the Knight shift will be the same. Since
$A_{d}(\mathbf{R}_{i})$ falls off with distance, one would observe
a spread of Knight shifts and the spread would continuously increase
in the same sense as the impurity susceptibility as the temperature
is lowered. Ultimately, the susceptibility will saturate at the lowest
temperatures which would correspond to a maximum spread of the Knight
shifts. The same can be said for the relaxation rates (see (\ref{eq:T-par-perp-kondo})).
Such behaviour of the Knight shift has been reported in Cu:Fe alloys
\cite{alloul}.

Combining (\ref{eq:T-par-perp-kondo}) and (\ref{eq:knight-shift2})
we get \cite{narath,shiba}\begin{eqnarray}
K(\mathbf{R}_{i})^{2}T_{1}(\mathbf{R}_{i})T & = & \frac{(g_{s}\mu_{B})^{2}}{(g_{n}\mu_{n})^{2}}\frac{\hbar}{4\pi k_{B}}.\label{eq:korringa-kondo}\end{eqnarray}
 Eq.(\ref{eq:korringa-kondo}) has the form of Korringa relaxation.\cite{korringa}

\subsection{Relaxation due to conduction electron coupling}

Relaxation due to coupling to conduction electrons matters only for
those nuclei that are so far from the impurity that their RKKY coupling
to the impurity is weaker than their hyperfine coupling with the conduction
electrons. That happens when $k_{F}R_{i}\gg1.$ As the temperature
falls below the impurity Kondo temperature, there is an enhancement
in the density of states at the Fermi energy: \cite{nozieres}\begin{eqnarray}
\tilde{\rho}(\epsilon_{F})=\rho(\epsilon_{F})[1+C(T_{F}/T_{K})1/N],\quad T\ll T_{K}\label{eq:rho-enhancement}\end{eqnarray}
 where the tilde denotes the Kondo-enhanced density of states at the
Fermi energy, $C$ is a constant of order one, and $N$ is the number
of electrons in the QPC. This leads to an enhancement of the relaxation
rate \cite{roshen1} given in (\ref{eq:cond-el-highT}): \begin{eqnarray}
T_{1}^{-1} & = & \frac{\pi(k_{B}T)}{\hbar}(A_{s}\tilde{\rho}(\epsilon_{F}))^{2},\quad T\ll T_{K}.\label{eq:enhanced-rho-T1}\end{eqnarray}
 Thus we can summarise, \begin{eqnarray}
\frac{T_{1}^{-1}|_{T\ll T_{K}}}{T_{1}^{-1}|_{T\gg T_{K}}} & = & \frac{\tilde{\rho}(\epsilon_{F})^{2}}{\rho(\epsilon_{F})^{2}}\approx1+2C(T_{F}/T_{K})1/N.\label{eq:T1-ratios}\end{eqnarray}
 We should perhaps use this enhanced density of states even for the
case of relaxation through coupling to the impurity spin below the
Kondo temperature.

Application of a magnetic field will tend to decrease the density
of states towards the high temperature value. In the Kondo regime,
the impurity susceptibility is proportional to the density of states
of the conduction electrons. From the known Bethe ansatz solution
for the impurity magnetisation, we can extract the magnetic field
dependence of the density of states: \cite{roshen2}\begin{eqnarray}
\tilde{\rho}(\epsilon_{F},H_{0}) & \approx & \rho(\epsilon_{F})\left[1+\frac{CT_{F}}{NT_{K}}\left(1-C'\left(\frac{g_{s}\mu_{B}H_{0}}{k_{B}T_{K}}\right)^{2}\right)\right],\label{eq:rho-H}\end{eqnarray}
 where $C'$ is a constant of order one.

\section{Crossover between high and low temperature regimes \label{sec:Crossover}}

We have two independent parameters demarcating low and high temperature
behaviour: $\omega_{e}T_{e2}$ and $T/T_{K}.$ So we need to discuss
further the meaning of low and high temperature regimes.

The Kondo temperature is approximately $T_{K}\approx\epsilon_{F}e^{-1/J\rho(\epsilon_{F})},$
where $J\rho(\epsilon_{F})$ is the unrenormalised, i.e., bare, Kondo
coupling. Given that $\epsilon_{F}\approx20{\rm K},$ we cannot have
too small a value for $J\rho(\epsilon_{F})$ if we are to have any
hope of probing the behaviour on either side of the Kondo temperature.
Even for $J\rho(\epsilon_{F})=0.1,$ we would get a very small $T_{K}\approx10^{-3}{\rm K}.$
Let us therefore assume that the bare $J\rho(\epsilon_{F})\lesssim1.$

In our discussion of the behaviour above $T_{K},$ we had obtained
two regimes depending on the magnitude of $\omega_{e}T_{e2}.$ A small
value of $\omega_{e}T_{e2}$ corresponded to a high temperature. From
(\ref{eq:Te-high-temp}), we can see that the criterion for high temperature
behaviour (\ref{eq:T-par-perp-highT}) is \begin{eqnarray}
T & \gg & T_{\mbox{high}}=\frac{\hbar\omega_{e}}{\pi k_{B}(J\rho(\epsilon_{F}))^{2}}.\label{eq:high-T-criterion}\end{eqnarray}
 This is not too different from the temperature corresponding to the
Zeeman splitting of the localised electron given our expectations
regarding the value of $J\rho(\epsilon_{F}).$ Now the Kondo temperature
can either be larger or smaller than $T_{\mbox{high}}.$

Suppose $T_{K}\ll T_{\mbox{high}}.$ Then in principle we have \emph{three}
regimes: $T\gg T_{\mbox{high}},$ $T_{K}\ll T\ll T_{\mbox{high}},$
and $T\ll T_{K}.$ In the high temperature regime, $T\gg T_{\mbox{high}},$
we will observe a non-Korringa relaxation, (\ref{eq:T-par-perp-highT}),
due to coupling with the impurity spin.

Note that the condition $T_{K}\ll T_{\mbox{high}}$ corresponds to
$T_{K}\ll g_{s}\mu_{B}H_{0}/k_{B}.$ However all our discussion of
$T\ll T_{K}$ assumed that the Zeeman splitting of the impurity was
less than the Kondo temperature. We should not use those results for
$T\ll T_{K}.$ In fact, the large Zeeman field suppresses the {}``Fermi
liquid'' regime of the Kondo model. Thus \emph{there is no Kondo
regime} for $T_{K}\ll T_{\mbox{high}}.$ There are just two regimes
separated by $T_{\mbox{high}},$ and the relaxation rates in these
two regimes are given by (\ref{eq:T-par-perp-highT}), (\ref{eq:T-perp-lowT})
and (\ref{eq:T-par-lowT}). The maximum relaxation rate occurs around
$T_{\mbox{high}}$ where $\omega_{e}T_{e2}\approx1.$

Suppose $T_{K}\gg T_{\mbox{high}}.$ If the impurity Zeeman splitting
is small, then this is the likely scenario. In that case we should
redefine our high temperature regime to mean $T\gg T_{K}.$ Owing
to the qualitative change in the susceptibility and other properties
at $T<T_{K},$ we must not use (when $T<T_{K}$) (\ref{eq:T-par-perp-highT}),
(\ref{eq:T-perp-lowT}) and (\ref{eq:T-par-lowT}) which were derived
assuming a Curie susceptibility for the impurity spin and the bare
value of the dimensionless Kondo coupling. Such assumptions are correct
only when $T\gg T_{K}.$ In the low temperature regime, $T\ll T_{K},$
the relaxation will be given by (\ref{eq:T-par-perp-kondo}). In the
region of $T=T_{K},$ the ratio of the relaxation rate on the high
temperature side to the Kondo side is of the order of $1/(J\rho(\epsilon_{F}))^{2}.$
Since the coupling constant $J\rho(\epsilon_{F})$ diverges below
$T=T_{K},$ the Kondo relaxation rate will dominate near $T=T_{K}$
and below. As the temperature is decreased starting from the high
temperature side, one would observe a steady enhancement of the relaxation
rate (obeying the $1/T$ law) up to $T\sim T_{K},$ followed by a
linear-$T$ decrease according to (\ref{eq:T-par-perp-kondo}). (Maximum
relaxation rate at $T\approx T_{K}.$)

Further confirmation of the Kondo effect can be made by measuring
the temperature dependence of the Knight shift as shown in (\ref{eq:knight-shift2}).
If the temperature dependence of the Knight shift is the same as that
of the Kondo impurity susceptibility both above and below the Kondo
temperature, then the Kondo effect will be confirmed.

\section{Relaxation by impurity coupling and conduction electron
coupling \label{sec:Discussion1}}

Let us compare the relative magnitudes of relaxation by coupling to
the impurity spin and to conduction electrons. Consider the low temperature
regime, $T\ll T_{K},$ and a small magnetic field such that $T_{K}\gg g_{s}\mu_{B}H_{0}/k_{B}.$
Thus we need to compare the relaxation rates in (\ref{eq:T-par-perp-kondo})
and (\ref{eq:enhanced-rho-T1}). First consider nuclei inside the
region $V_{0}$ about the impurity. In this region, we have mentioned
earlier that $A_{d}(\mathbf{R}_{i})\approx8A_{s}/(w_{x}w_{y}w_{z}).$
It is easy to see that the ratio of the relaxation rates through coupling
with the impurity and with the conduction electrons is of the order
of $(A_{d}(\mathbf{R}_{i})\chi_{\mbox{imp}}/(g_{s}\mu_{B})^{2})^{2}/(A_{s}\rho(\epsilon_{F}))^{2}\sim(4\pi\hbar^{2}k_{F}/mw_{x}k_{B}T_{K})^{2},$
where we used $\rho(\epsilon_{F})=4m/(2\pi\hbar^{2}k_{F}w_{y}w_{z}).$
Estimating $2\pi/w_{x}\sim k_{F},$ the ratio works out to $\sim(4\epsilon_{F}/T_{K})^{2}\gg1.$
Therefore in the region $V_{0}$ around the impurity electron, nuclear
relaxation is primarily through coupling with this electron. Outside
$V_{0},$ the impurity RKKY coupling decreases as $1/(k_{F}R_{i}).$
The distance at which relaxation by conduction electrons becomes comparable
depends on the strength of $J\rho(\epsilon_{F}).$ We have argued
before that we need $J\rho(\epsilon_{F})\lesssim1$ in order to have
any chance of measuring on both sides of the Kondo temperature with
the usual apparatus. Thus the RKKY interaction is smaller than $A_{s}$
by a factor of $1/(k_{F}R_{i}).$ Therefore the distance beyond which
relaxation is mostly by conduction electron coupling corresponds to
$(4\epsilon_{F}/k_{B}T_{K})^{2}/(k_{F}R_{i})^{2}<1,$ or $R_{i}>4\epsilon_{F}/(k_{B}T_{K}k_{F}).$

The Kondo impurity, if present, will be easier to detect through its
direct or RKKY exchange coupling with the nuclear spins for three
reasons. First, we have already seen above that the higher susceptibility
of the impurity compared to the conduction electron susceptibility
for $T\ll T_{F}$ leads to a stronger nuclear relaxation rate. Second,
the temperature dependence of the nuclear relaxation in the former
case does not follow the Korringa law at high temperatures. Third,
the Knight shift will broaden as the temperature is lowered, and the
temperature dependence of the broadening will be directly proportional
to the Kondo impurity susceptibility (which is well-known). All cases
we discussed obey the Korringa law at temperatures below the Kondo
temperature.

We have not discussed the role of possible electron-electron interaction.
Electron interaction will affect both the density of states as well
as the impurity susceptibility. Proximity to a ferromagnetic instability
of the conduction electrons will enhance the impurity susceptibility
(through enhancement of the electron gyromagnetic ratio) which will
tend to increase the relaxation rate. However one needs to keep in
mind any interaction effects on the density of states. In the absence
of the Kondo impurity, Moriya has shown that the nuclear relaxation
rate is enhanced by electron-electron repulsion \cite{moriya}.

\section{Relaxation by nuclear spin diffusion\label{sec:discussion2}}

In our treatment we have so far ignored internuclear dipolar interactions
that will cause internuclear flip-flops and nonconserving nuclear
spin flips. In GaAs, the intrinsic nuclear relaxation times $T_{1}$
and $T_{2}$ can roughly estimated to be of the order of $\hbar/\epsilon_{dd}\sim10^{-4}{\rm s},$
where $\epsilon_{dd}$ is the magnetic dipolar interaction of neighbouring
nuclei corresponding to a field of about $1{\rm mT}$ acting on the
nuclei. In non-zero fields, however, $T_{1}$ can be larger by several
orders of magnitude as for example has been observed \cite{mcneil}
in GaAs where $T_{1}\sim10^{3}{\rm s}$ at fields of about $140{\rm mT}.$
In the following discussion we assume that a field of several millitesla
is present so that nonconserving spin flips due to internuclear interaction
may be ignored.

In addition to nonconserving spin flips, one also has internuclear
spin flip-flop processes. The latter give rise to nuclear spin diffusion
(NSD) and occur even in the presence of an external magnetic field.
NSD effects in quantum dots are a topic of much recent study owing
to their importance for nuclear spin polarisation based qubits. A
thorough analysis of NSD is not attempted here given the incomplete
understanding in the literature of the same on quantum dots. Instead
we discuss qualitatively the conditions under which NSD effects can
be important in our case, and how this may be suppressed to allow
the electronic relaxation mechanisms to have a greater effect on the
QPC conductance. A simple model for studying the spatial dependence
and temporal decay of the nuclear polarisation is \begin{eqnarray}
\frac{\partial M}{\partial t} & = & D\nabla^{2}M-\frac{M-M_{0}}{T_{1}(\mathbf{r})},\label{eq:diffusion-eq}\end{eqnarray}
where $D$ is the nuclear spin diffusion constant and $M_{0}$ is
the steady state nuclear polarisation in the given external magnetic
field. The nuclear spin diffusion constant is related to the decoherence
time $T_{2}$ for the nuclear polarisation; for a cubic lattice such
as GaAs \cite{khutshishvili,bloembergen}, \begin{eqnarray}
D & \approx & \frac{a^{2}}{30T_{2}},\label{eq:spin-diffusion}\end{eqnarray}
 where $a$ is the nearest distance between nuclei of the same species.
In pure, \emph{bulk} GaAs, the internuclear flip-flop processes set
an upper limit to $T_{2}\sim\hbar/\epsilon_{dd}\sim10^{-4}{\rm s},$
which gives us $D_{bulk}\sim10^{-13}{\rm cm}^{2}/{\rm s}.$ Experimentally
observed values of the nuclear spin diffusion constant in bulk GaAs
due to internuclear dipolar interactions are in agreement with this
rough estimate \cite{paget}. 

In a quantum dot with a localised impurity electron, the spatial variation
of the localised electron wavefunction leads to a spatially varying
hyperfine contact interaction. This affects both the relaxation and
spatial distribution of the nuclear polarisation. First, the spatial
variation of the hyperfine interaction in the quantum dot has been
shown \cite{deng} to cause a suppression of the diffusion constant
$D_{dot}$ in the dot by a factor of the order of $10$ compared to
$D_{bulk}$ because nuclear flip-flop transitions in this case do
not conserve energy. Experimentally, the NSD constant in quantum dots
has also been reported to be small compared to the bulk value \cite{bayot,stephens}.
Second, during the build-up of the nuclear polarisation, the inhomogeneity
of the hyperfine interaction translates into an inhomogeneous nuclear
polarisation, with a maximum near the centre of the dot, and rapid
decay outside the dot. Due to the presence of the diffusion term,
the solution of Eq.(\ref{eq:diffusion-eq}) with a nonuniform initial
distribution of nuclear polarisation does not in general decay exponentially
with time \cite{tifrea}. Exponential decay can however take place
if the diffusion energy in Eq.(\ref{eq:diffusion-eq}) is smaller
than $\hbar/T_{1}.$ We estimate the nuclear diffusion rate $1/T_{1}^{sd}$
to be the order of $D_{dot}/l_{min}^{2},$ where $l_{min}$ is the
smallest dimension of the QPC along which nuclear spins may diffuse.
In our case, $l_{min}=w_{y}=5{\rm nm},$ and conservatively using
for $D_{dot}$ the bulk diffusion value $D_{bulk}$ for GaAs, we find
$T_{1}^{sd}\approx0.4{\rm s}.$ If we take into account the suppression
of the diffusion constant in the quantum dot because of an inhomogeneous
hyperfine interaction \cite{deng}, we will have $T_{1}^{sd}\sim4{\rm s}$
for $D_{dot}\sim0.1D_{bulk}.$ In recent measurements on quantum dots
\cite{maletinsky}, enhancement of the nuclear relaxation time by
a factor of nearly two orders of magnitude (to nearly $100{\rm s}$)
has been reported at fields more than $1{\rm mT}.$ Another way to
increase the NSD time is by designing the 2DEG such that we have AlGaAs
layers on either side of the 2DEG, instead of on one side as we have
considered here. NSD is suppressed in a direction perpendicular to
the 2DEG because of the change of material from GaAs to AlGaAs as
well as disorder in AlGaAs \cite{malinowski}. In such a redesigned
QPC, we should regard the transverse width $w_{z}=20{\rm nm}$ as
$l_{min},$ and that will give $T_{1}^{sd}\sim6.4{\rm s}$ even if
inhomogeneous hyperfine interaction effects are not taken into account,
and $T_{1}^{sd}\sim65{\rm s}$ if this is taken into account. We note
that in experiments on quantum dots in Ref. \cite{ono}, $T_{1}^{sd}$
has been estimated to be as long as $200{\rm s}.$ 

To compare with the nuclear relaxation rates in the Kondo scenario
which is the subject of this paper, we have for the QPC $A_{d}\approx5.8\times10^{-29}{\rm J}$
per nucleus and we associate the experimental energy scale determining
the conductance with the Kondo temperature: $T_{K}\approx1{\rm K}.$
For a QPC defined in a GaAs 2DEG with conduction electron density
$10^{11}{\rm cm}^{-2},$ the 1D Fermi energy $\epsilon_{F}$ ($m=0.067m_{e}$)
in the lowest sub-band is about $20{\rm K};$ and using $T_{K}\approx\epsilon_{F}e^{-1/J\rho(\epsilon_{F})},$
we estimate the bare (high temperature) value of $J\rho(\epsilon_{F})\approx0.35.$
In the {}``high'' temperature region ($T>T_{K}$), say $T=2{\rm K},$
Eq.(\ref{eq:T-par-perp-highT}) then gives the relaxation time due
to coupling to the impurity electron as $T_{1}^{imp}\approx0.1{\rm s}.$
This is comparable with our most conservative estimate above for the
relaxation time due to nuclear spin diffusion, while if we take into
account the suppression of NSD due to inhomogeneous hyperfine interaction,
and/or design the 2DEG to suppress diffusion perpendicular to the
2DEG, NSD effects are much smaller and may be ignored in a first treatment.
The relaxation time using the above parameters due to coupling to
conduction electrons as estimated from Eq.(\ref{eq:cond-el-highT})
is $T_{1}^{cond-el}\approx5{\rm s},$ which is also long compared
to relaxation by coupling to the paramagnetic impurity. In the {}``low''
temperature region ($T<T_{K}$), the relaxation time associated with
coupling to the paramagnetic impurity as given by Eq.(\ref{eq:T-par-perp-kondo})
(using $\chi_{\mbox{imp}}\approx(g_{s}\mu_{B})^{2}/k_{B}T_{K}$) is
$T_{1}^{imp}\approx3.5\times10^{-2}{\rm s}$ at $T=0.5{\rm K},$ which
is much shorter than the relaxation times $T_{1}^{sd}\sim10{\rm s}$
and $T_{1}^{cond-el}$(at this temperature $T_{1}^{cond-el}\approx20{\rm s}$)
respectively due to NSD and coupling to conduction electrons. The
latter two effects are therefore safely ignored in the QPC, except
at very low temperatures when NSD may dominate because it does not
vanish at $T=0.$ Away from the centre of the QPC, relaxation by coupling
to the paramagnetic impurities and coupling to conduction electrons
become comparable. We estimate this distance from the discussion in
Sec. \ref{sec:Discussion1} to be $R_{i}=(4\epsilon_{F}/k_{B}T_{K}k_{F})\approx1.6\mu{\rm m},$
which is of the order of the length of the QPC. Thus outside the QPC,
relaxation by coupling to conduction electrons is also important.
It is easily seen that the same is also true for NSD. Nevertheless,
since the conductance is very sensitive to the Overhauser field in
the QPC and not to the Overhauser field in the 2DEG, we conclude that
to a first approximation, $T_{1}$ obtained from the conductance of
the QPC is dominated by the coupling to the paramagnetic impurity
compared to nuclear spin diffusion and coupling to conduction electrons. 

To summarise, nuclear spin diffusion effects may be ignored in our
analysis if the experiments are performed in fields of several millitesla,
and the temperature is high enough such that the nuclear diffusion
time $l_{min}^{2}/D$ is much longer than the relaxation time $T_{1}$
from electronic processes. A more accurate treatment of NSD effects
is needed at very low temperatures and for long QPCs. 

V.T. thanks the support of TIFR and a DST Ramanujan Fellowship {[}sanction
no. 100/IFD/154/2007-08]. N.R.C. acknowledges support by EPSRC grant
GR/S61263/01.

\end{document}